# Thermodynamic driving force in the formation of hexagonal-diamond Si and Ge nanowires

E. Scalise[1] (*), A. Sarikov[1,2], L. Barbisan[1], A. Marzegalli[3], D. B. Migas[4,5], F. Montalenti[1] and L. Miglio[1]

[1] *L-NESS and Department of Materials Science, Università degli Studi di Milano-Bicocca, via Cozzi 55, 20125, Milano, Italy*
[2] *V. Lashkarev Institute of Semiconductor Physics, National Academy of Sciences of Ukraine, 45 Nauki avenue, 03028, Kyiv, Ukraine*
[3] *L-NESS and Department of Physics, Politecnico di Milano, via Anzani 42, 22100, Como, Italy*
[4] *Department of micro- and nanoelectronics, Belarusian State University of Informatics and Radioelectronics, P. Browka 6, 220013, Minsk, Belarus*
[5] *National Research Nuclear University MEPhI (Moscow Engineering Physics Institute), Kashirskoe shosse 31, 115409 Moscow, Russi*
*Corresponding author: emilio.scalise@unimib.it

**ABSTRACT**

The metastable hexagonal-diamond phase of Si and Ge (and of SiGe alloys) displays superior optical properties with respect to the cubic-diamond one. The latter is the most stable and popular one: growing hexagonal-diamond Si or Ge without working at extreme conditions proved not to be trivial. Recently, however, the possibility of growing hexagonal-diamond group-IV nanowires has been demonstrated, attracting attention on such systems. Based on first-principle calculations we show that the surface energy of the typical facets exposed in Si and Ge nanowires is lower in the hexagonal-diamond phase than in cubic ones. By exploiting a synergic approach based also on a recent state-of-the-art interatomic potential and on a simple geometrical model, we investigate the relative stability of nanowires in the two phases up to few tens of nm in radius, highlighting the surface-related driving force and discussing its relevance in recent experiments.

We also explore the stability of Si and Ge core-shell nanowires with hexagonal cores (made of GaP for Si nanowires, of GaAs for Ge nanowires). In this case, the stability of the hexagonal shell over the cubic one is also favored by the energy cost associated with the interface linking the two phases. Interestingly, our calculations indicate a critical radius of the hexagonal shell much lower than the one reported in recent experiments, indicating the presence of a large kinetic barrier allowing for the enlargement of the wire in a metastable phase.

# 1   Introduction

Silicon (Si) is the second most abundant element in the Earth's crust, after oxygen. The natural abundance, besides minimal toxicity, the feasibility of large single-crystals and its very stable native oxide, contributed to making Si the most important and used semiconducting material in electronics and photovoltaics. The maturity of silicon technology-enabled ubiquitous integrated circuits (ICs) has driven the advances in micro- and nanoelectronics. However, the poor optical properties of cubic diamond (*3C, $Fd\bar{3}m$*) crystalline silicon, originating from the indirect nature of its electronic band gap, constitute a major weakness for photonic and optoelectronic applications and hinder the integration of the optically active components in a monolithic Si substrate.

The recent successes in the synthesis of Si crystals with the hexagonal diamond (*2H, $P6_3/mmc$*) crystalline structure [1–10] paved the way for the realization of on-chip light sources based on the hexagonal allotropes of group IV elements and highly compatible with the current Si technology. Although the *2H* crystal phase retains the indirect nature of the *3C*-Si band gap [10], theoretical predictions [11], confirmed by recent photoluminescence investigations, show a direct band-gap for 2H-Ge, which is preserved by alloying Ge with up to 30% Si [11,12].

All the recent successful attempts in obtaining the *2H*-Si and Ge have exploited nanostructures and especially the nanowires (NWs), allowing to overcome the need for extreme pressure conditions of classical phase-transition experiment previously employed to obtain the *2H* crystal phase [13]. Two main methods have been recently used to produce Si (or Ge) NWs with the *2H* crystal phase. Initially, the Si NWs have been mainly synthesized by vapour-liquid-solid (VLS) mechanism, using catalytic metal nanoparticles [6,7,14]. The *2H*-Si NWs obtained by VLS are typically not very uniform, with the *3C* crystal phase still dominating, or they have very small diameters [15]. Lately, while the VLS method has been still used to obtain GaP or GaAs NWs in the wurtzite phase (i.e. hexagonal structure), the Si(Ge) shell on top has been obtained by Chemical Vapor Deposition (CVD). Both the Si [1] and SiGe [2,12] shells were unambiguously proven to have a high-quality *2H*-crystalline structure. All these advances [16] evidence how the allotropes of the group IV compounds may be an optimum solution to the need for a direct-gap material compatible with the Si technology. Thus, the identification of the driving force in the crystallization of *2H*-Si and Ge crystals is a potential breakthrough in this field.

A key feature of low dimensional nanostructures, including NWs, is the high surface to (bulk) volume ratio. Indeed, the surface energy plays a crucial role in the formation and stability of the NWs, particularly in case of small radii and also at the initial stage of the epitaxial growth of core/shell systems, being the surfaces still dominant. Indeed, the NW diameter and its specific sidewall facets have been often pointed out as key factors determining the crystal phase of III-V NWs [17,18]. Moreover, in the core/shell NW structures, the interface between the core and shell may strongly influence the crystal phase of the shell, with the core getting a template effect on the shell.

In this work, we investigate the impact of the interfaces and surfaces at the sidewall facets of the NWs on their thermodynamic stability. By exploiting the surface and interface energies of different crystals of both the *3C* and *2H* crystal phases, we show that it is possible to estimate a critical radius of the NWs, up to which the *2H* crystal phase is thermodynamically more stable than the *3C* one. This is further supported by simulating the whole Si and Ge NWs, both with the *2H* and *3C* structure and including surface reconstructions. For small radii of the NWs, the total formation energy of the NWs is lower for the *2H* than the *3C* NWs, but this is different when NWs with larger radii are simulated. Our work demonstrates that there is a thermodynamic driving force of the surfaces in the formation of Si and Ge NWs in the metastable *2H* phase. This force is likely significant in the synthesis of small *2H*-Si NWs by VLS methods [15], and also essential in the growth of core/shell *2H*-Si and Ge (and SiGe) NWs [12], where the template effect of the core is also key.

## 2      Results and discussion

### 2.1 *2H*-Si and Ge surfaces

One of the most interesting features of low-dimensional nanostructures, which is indeed related to their high surface-to-volume ratio, is the possibility to tune their optoelectronic properties by engineering their surfaces and interfaces [19,20]. This peculiarity is often fundamental even in the synthesis of nanostructures, as for instance in the case of colloidal nanocrystals [21,22]. The thermodynamics of surfaces is also very important in bulk semiconductors, both in determining the reconstructions that are formed on the cleaved surfaces of the bulk crystals but also conferring the equilibrium shape to the crystals, obeying the famous Wulff construction [23]. Being motivated by all these considerations, we study the surface energies of both the *3C* and *2H*-Si and Ge crystals. In fact, a marked difference between the surface energies of the two polytypes may lead to a significant contribution to the total energy of the NWs, compensating the lower cohesive energy of the *2H* crystal phase and, crucially, providing a thermodynamic driving force for the formation of the *2H*-Si and Ge NWs. These considerations were at the base of two preliminary theoretical works by two independent groups [24,25] performed much earlier than the recent experimental successes in this field, but still not explicating them [15].

In Table 1 the surface energies of the two lowest-index surfaces perpendicular to the [111] or [0001] direction are listed for the *3C* and *2H*-Si (or Ge), respectively. In fact, the [0001] direction is the orientation of the *2H*-Si and Ge NWs recently obtained by the core/shell growth method [1,12,26,27] and showing the best crystal quality among the recent experimental attempts to synthetize hexagonal diamond Si and Ge crystals. Note also that [0001] oriented *2H*-Si NWs unambiguously appear in recent VLS growth experiments of very small-radius Si NWs [15].

The (ABAB…) stacking sequence of the hexagonal diamond structure along the [0001] direction corresponds to the (ABCABC…) stacking sequence of cubic one along the [111] direction. Hence, the cubic (111) surface can be placed right on top of the (0001) basal plane of the 2H structure, forming a coherent interface between the hexagonal and the cubic phase. Thus, the [111] direction of the cubic crystal is the most obvious direction to be compared to the [0001] hexagonal one when investigating the energetic competition of the two crystal phases in these nanowires. The [111] direction is also the orientation of the cubic nanowires exploited to obtain a strain-induced phase transformation from *3C* to *2H*-Ge [3,28], and it is also the most common direction of larger-radius NWs grown by VLS method on Si or Ge(111) substrates [29,30].

**Table 1** Surface energy [meV/Å$^2$] for different orientations of the 3C and 2H Si and Ge. Values with * are obtained by using the Csányi's potential as discussed in the method section.

|       | (111)   | (110)   | (112)   |
|-------|---------|---------|---------|
| 3C-Si | 80.89   | 94.19   | 87.37   |
|       | 81.67*  | 94.71*  | 91.53*  |
| 3C-Ge | 57.7    | 60.54   | 58.24   |

|       | (0001)  | (11$\bar{2}$0) | (10$\bar{1}$0) |
|-------|---------|----------------|----------------|
| 2H-Si | 78.56   | 79.09          | 72.55          |
|       | 79.34*  | 81.79*         | 76.38*         |
| 2H-Ge | 53.34   | 50.84          | 47.59          |

The literature on the cubic Si and Ge surfaces is very rich and particularly the low-index surfaces, such as the (001), the (110) and the (111) ones, are quite well understood [31–33]. The higher index surfaces, especially the (112) have received less consideration, mainly because it was often considered intrinsically unstable, suggesting that it decomposes into (111) and other high-index facets [34,35] at the termination of bulk-like samples. Whether the (112) sidewalls, which are clearly discernible in VLS grown Si NWs [36], are also composed by sawtooth structures including the (111) facets, or they have the nominal (112) orientation, is still not completely understood. However, this is not very central to the following discussion, since first-principles calculations evidenced that [111] Si and Ge NWs delimited by (112) crystal facets are thermodynamically stable [37,38]. Moreover, these calculations showed negligible differences between the formation energy of *3C*-Si and Ge NWs with different morphologies. Indeed, it is likely that the formation of (111) facets is favored over the (112) facets, but then the other higher index facets and their boundaries with the (111) facets, necessary to form a sidewall with a global (112) orientation, will raise the total energy of sawtooth structures as compared to sidewalls formed only by nominal (112) facets. Thus, for the sake of comparison between the formation energy of cubic and hexagonal NWs, considering (112) facets depicts the best-case scenario for the cubic NWs oriented along the [111] direction.

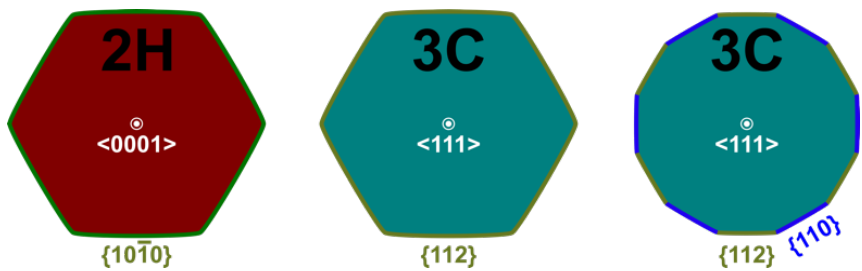

**Figure 1** Sketch of the nanowires shape with the different orientations. From left to right: a *2H* NW oriented along the <0001> direction and having an hexagonal shape with 6 {10$\bar{1}$0} facets; a <111> oriented *3C* NW with hexagonal shape and 6 {112} facets; a dodecagonal *3C* NW with 6 {112} and 6 {110} facets.

Very few studies report on the Si or Ge (112) surface [39–41], but they describe very well the two possible reconstructions, having 2×1 or 1×1 periodicity. In the former case, a dimer is formed between two top Si atoms, reducing the 4 dangling bonds per unit cell of the as-cut surface to 3. In the 1×1 reconstruction, one of the top atoms of the as-cut surface is removed and by forming also a dimer, two dangling bonds per 1×1 cell are still present. Due to the missing atom, the top bonds of this 1×1 reconstruction are very stretched. Nevertheless, most of the previous density functional theory (DFT) calculations, which were based on the local density approximation (LDA) [42], indicate the 1×1 reconstruction as the most stable. On the contrary, our DFT calculations, which are based on the generalized gradient approximation (GGA) (see the method section) predict the 2×1 reconstruction as the most stable, in agreement with results reported in a previous theoretical study of the morphology of <111> oriented 3C-Si and Ge NWs [37]. The calculated surface energy value is reported in Table 1, both for 3C-Si and Ge, and it is compared to the value calculated for the (110) surface. In fact, this is another crystal orientation that has been identified in the delimiting facets of 3C-Si or Ge NWs grown in the [111] direction, particularly in case of the dodecagonal morphology of the NWs [37,43]. Indeed, the surface energy of the Ge(110) is very close to that calculated for the Ge(112), and an equilibrium shape of the [111] NWs showing 12 facets with altering orientations is thermodynamically consistent. For Si, the calculated values for these two cubic surfaces manifest a noticeable difference that may be interpreted as a thermodynamic tendency to a hexagonal morphology, with dominating (112) surfaces. An exhaustive analysis of the impact of the surface energy on the shape of the NWs should consider also kinetic effects and would need more experimental evidence, but is out of the scope of this work. For the current study, either the dodecagonal shape with both (112) and (110) surfaces or the hexagonal morphology with only (112) would give very similar total energy of the NWs (see Fig. 1). In fact, in the former case the two classes of surfaces must have the same surface energy to coexist at the thermodynamic equilibrium. On the contrary, if the (112) surface energy is much lower, as for the Si case, then one has to consider the hexagonal shape, obviously minimizing the formation energy of the system. The calculated values for the (110) surfaces are in agreement with previous reported value [31], but slightly lower, due to the different DFT approximation employed (LDA vs. GGA). Similar differences with previous reported values in the literature [31,32] are also obtained for the most common Si and Ge surfaces (111), which have been also included in Table 1. Note that we have not modeled the 7×7 Si reconstruction due to the much larger supercell required, as compared to the simulated 2×1 reconstruction, but it is well known [32] that the former reconstruction is more stable by only a few meV/Å$^2$. Note also that in Table 1 are reported together with the Si surface energy values calculated by DFT, also the corresponding values obtained by exploiting a state-of-the-art interatomic potential for Si, recently developed by Csányi's group and based on machine learning methods [44]. The importance of this approach will be clear in the next section, where simulations of large-radius NWs, not accessible by DFT calculations, are presented. Finally, the surface energies for the *2H* crystallographic planes are also calculated. To the best of our knowledge no surface energy values are reported in the literature for the *2H*-Si and Ge. We show in Fig. 2 the (10$\bar{1}$0) and (11$\bar{2}$0) surfaces after atomic relaxation. Well-defined surface reconstructions dominated by the puckering of the top atoms are evident. Particularly for the (10$\bar{1}$0) surface, one may clearly note two dangling bonds per unit cell. The puckering of the top atoms with the dangling bonds allows their bond angles to vary substantially from the ideal angle of the sp$^3$ hybridization (close to 109°). Two very different bond angles are formed: one increasing above 120° and thus indicating a sp$^2$-like hybridization, with the dangling bond assuming a p-orbital character; on the contrary, the bond angle of the other atom decreases below 100° and suggests that the dangling bonds have a more s-like behavior in this case. This mechanism of reducing the energy of the dangling bonds has been observed also for surface reconstructions [45] of the *3C*-Si and particularly for the (112) reconstruction [41]. Contrary to the (112) surface, the analyzed surfaces of the *2H* phase do not form any new dimers, and the number of dangling bonds is not reduced upon reconstructions. Still, the density of dangling bonds (db) is estimated to be 0.0819 db/Å$^2$ for the Si(112) 2×1 reconstruction and 0.0817 db/Å$^2$ for the Si(10$\bar{1}$0) surface. Besides, an analysis of the bond lengths of the first and second atomic layer also reveals that the deviation from bulk bond length (weighted by the surface area) is higher for the (112) surface as compared to the (10$\bar{1}$0) surface (7.45·10$^{-3}$/Å and 6.94·10$^{-3}$/Å, respectively, for the Si surfaces). Despite these marginal differences, the surface energy of the (10$\bar{1}$0) surface is substantially lower than of the (112) one, both for Si and Ge. More in general, the energy difference between the (111) and the (0001) surfaces is enlightening, due to their identical bilayer termination, so that the reconstructions of these two surfaces are expected to be the same, and have been both modeled as the 2x1 chain-configuration. For this reason, the (moderate) surface energy difference, in this particular case, is just a natural consequence of the different cohesive energy in the 3C and 2H crystals: a lower binding energy of the 2H structure provides a lower surface energy for the (0001), with respect to the (111). Still, the sensibly larger values of the surface energy difference for the other two orientations, i.e (10$\bar{1}$0) vs (112) and (11$\bar{2}$0) vs (110) for both Si and Ge, indicate that the way the hexagonal phase rearrange the surface structure is likely to be more efficient, as a comparative analysis of the radial and angular bond distributions qualitatively suggests (not shown here). Actually, a general tendency showing lower surface energies for the hexagonal as compared to the cubic crystal has been reported also for ZnS [46], though the identical cohesive energy of the two crystal phases. However, a more specific investigation could be useful to understand whether this general tendency is a direct consequence of the different stacking sequence, or it is more related to the specific material.

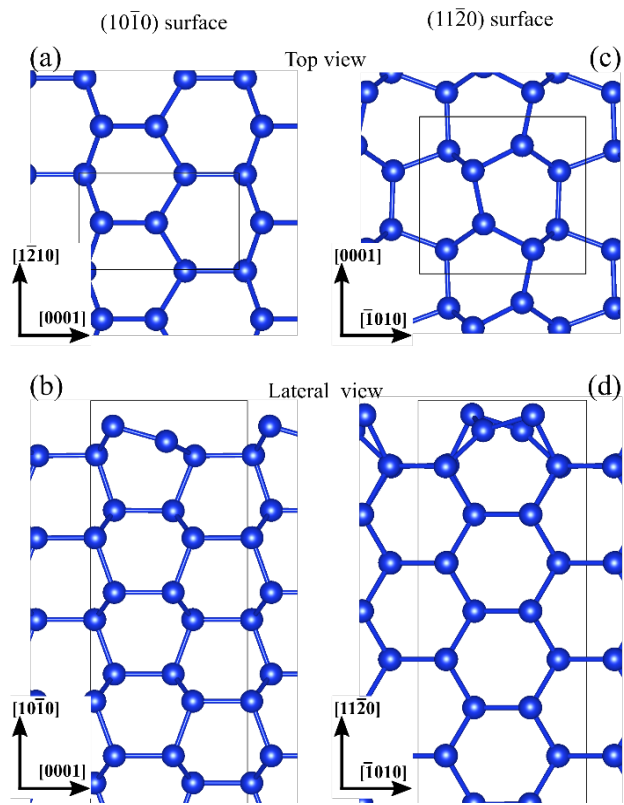

**Figure 2** The ($10\bar{1}0$) and ($11\bar{2}0$) surfaces in panel a-b and c-d, respectively. Both top (panels a and c) and side views (b and d) of the simulated slabs are shown.

In the next section we exploit the large difference between the $(10\bar{1}0)$ and $(112)$ surfaces to show that a crossing point between the formation energy of *3C* and *2H* NWs can be precisely estimated, clearly showing the thermodynamic stability of small-radius *2H*-Si and Ge NWs along the [0001] direction (or [111] if referred to the cubic crystal).

## 2.2 Diameter dependence of the stability of *2H*-Si and Ge NWs

Two different approaches are used to calculate and compare the total energy of the *2H* and *3C*-Si and Ge NWs. The first one exploits the surface energy values presented above in a direct manner, i. e. by inserting them in an analytical model of the formation energy of the NWs, summing up the surface energy of the sidewalls of hexagonal-shaped NWs to their bulk energy. This is expressed in the following Eqs. (1 and 2), showing the dependence of the formation energy (as referred to the *3C* bulk crystal) on the NW radius.

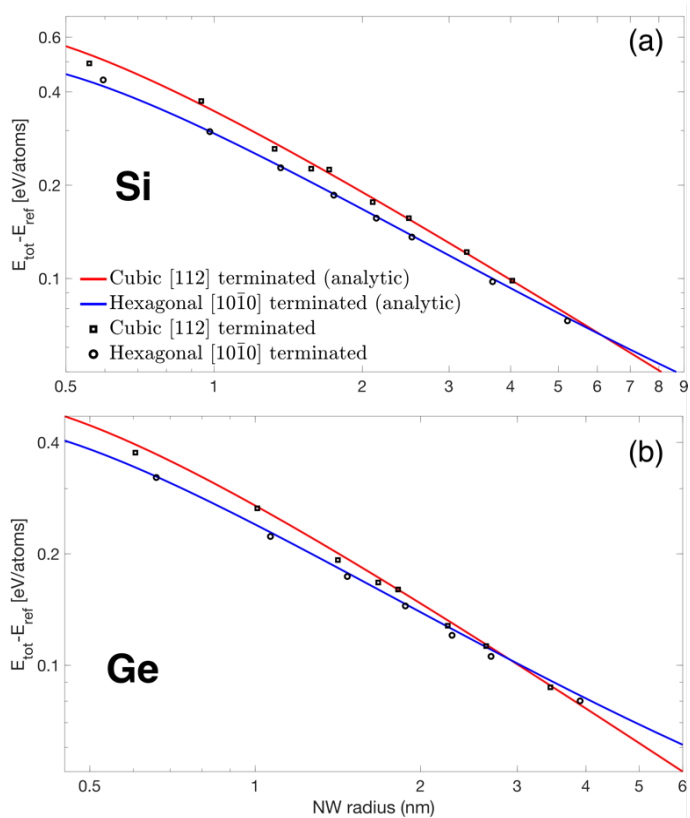

**Figure 3** Total energy of *2H* and *3C*-Si (a) and Ge (b) NWs relative to the 3C bulk crystal. Solid lines are the energy curves obtained by the analytical models described in the text, while the points represent the total energy explicitly obtained from the atomistic simulation of the NWs.

$$\Delta E_h = (\mu_h - \mu_c) + \frac{\frac{4}{\sqrt{3}} \gamma_{10\bar{1}0} \, Vol^h_{atm}}{r_{nw} - 0.16 \, nm} \qquad (1),$$

$$\Delta E_c = \frac{\frac{4}{\sqrt{3}} \gamma_{112} \, Vol^c_{atm}}{r_{nw} - 0.16 nm} \qquad (2).$$

The NW radius is indicated by $r_{nw}$, γ stands for the surface energy as investigated previously, while ($\mu_h$-$\mu_c$) is the difference between the chemical potentials of *2H* and *3C* bulk crystals, i.e. the formation energy of the *2H* phase. It is predicted to be about 11 meV (10 meV by Csányi's potential) for the *2H*-Si and 19 meV for the *2H*-Ge. The specific atomic volume of the *2H* and *3C* crystal phases is indicated by $Vol_{atm}$. The calculated formation energies of cubic and hexagonal NWs are plotted as functions of the NW radius, both for Si in Fig. 3(a) and for Ge in Fig. 3(b).

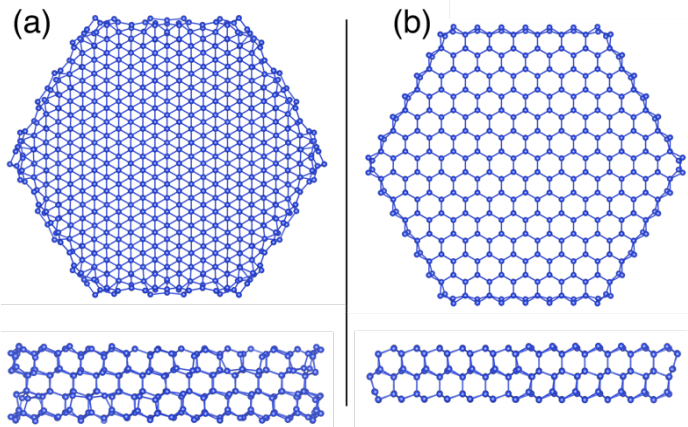

**Figure 4** Atomistic models of *3C* (a) and *2H*-Si (b) NWs. The top figures show the cross-section of the NWs, while the bottom ones are side views of the periodic structures used to model infinite long NWs. The *3C*-Si NW model contains 842 Si atoms, while the *2H* one has 588 atoms, and their radii are estimated to be about 2.8 nm and 2.7 nm, respectively.

Beside the analytic model, some points in the plots of Fig. 3 indicate the total energy of the NWs calculated with a second approach. In this case, the NWs are modelled by cutting a bulk crystal and creating hexagonal prisms as illustrated in Fig. 4. These structures are periodic in the [111] direction for the *3C* (see Fig. 4(a)) or [0001] for the *2H* crystals, hence infinite long NWs elongated in these directions are simulated. The sidewalls surfaces of the NWs belong to the two families of crystal planes considered previously, namely the {112} and the {10$\bar{1}$0} for the *3C* and *2H* crystal phase, respectively. Specific radii are chosen in order to have maximum one dangling bond per atom on these surfaces and few manipulations are also needed to facilitate the reconstruction of the lowest-energy surface after DFT atomic relaxation and to obtain sidewalls almost identical to the crystal surfaces discussed in the previous section.

The curves of the Eqs. 1 and 2 are in very good agreement with the total energy values that we explicitly calculated by the NW simulations and confirm the thermodynamic stability of small radius *2H* NWs. This also suggests that the edge contribution, which is not considered in the analytic model but in the explicit atomistic simulations, is not critical in this specific case, whereas it is typically significant in the formation energy of nanostructures.

A remarkable and novel result is the clear difference between the Si and Ge case. For the Si NWs, we estimate the crossover between the total energy of 2H and 3C Si at about 6.2 nm, while for Ge this occurs at about 3 nm. Thus, particularly for the Si case, our prediction of the maximum radius for thermodynamic stability of the *2H* NWs is quite larger than that considered in previous calculations [24,25] and even larger than the radius of *2H* NWs synthetized by VLS [15]. While for Ge the crossover predicted analytically is confirmed by the DFT energy trends obtained by the direct simulation of the NWs (see Fig. 3(b)), the large radius of the Si NWs makes a direct calculation of the NW total energy by DFT almost inaccessible. Then, we exploited classical simulation based on the Csányi's potential to estimate the total energy of Si NWs with radii much larger than those considered in the graph of Fig. 3(a).

In Fig. 5 the results obtained by using the Csányi's potential are plotted, and they include the total energy of NWs with radii approaching 18 nm. For these calculations, analogously to the previous DFT case, the analytical trends agree very well with the energy values calculated by direct modelling of the NWs. Note that the predicted crossover value is only slightly larger than the previous DFT value (about 7.1 nm). The results obtained by the Csányi's potential are very meaningful, because they provide a further and complementary proof of the moderately large radius for the thermodynamic stability of the *2H*-Si NWs. In fact, it has been possible to perform atomistic simulations of the NWs with sizes beyond that of the crossover in energy predicted analytically, providing for the first time a direct evidence of the crossover between the formation energy of *3C* or *2H*-Si. Besides, they are also a remarkable test-case for the Csányi's potential in itself.

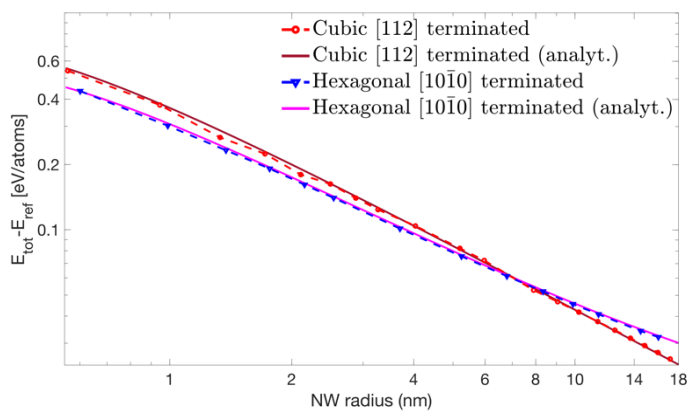

**Figure 5** Total energy of *2H* and *3C*-Si NWs relative to a bulk crystal. Solid lines plot the analytical models, while the dashed lines join the total energy points derived from calculations of the full NWs by the Csányi's potential.

Although other investigations are necessary to also elucidate the role of atomic species diffusing from the catalytic nanoparticles and of possible kinetic effects in the formation of *2H*-Si NWs by VLS [15,16], our study indubitably shows that there is a thermodynamic driving force in the formation of *2H*-Si and Ge NWs due to surface effects, up to radii of about 6 nm for Si or 3 nm for Ge NWs.

**2.3 Core/shell interface and template effect**

In core/shell NWs [1,2,12,26,27], the process leading to the formation of the *2H*-Si or Ge shells on top of wurtzite III-V (typically GaP or GaAs, respectively ) cores, differs significantly from the VLS growth. Here, an important role is expected to be played by the template effect of the cores [16]. The role of the interface for the thermodynamic stability of core/shell *2H*-Si or Ge NWs is now investigated and the relation between the *2H* phase stability and the NW radius, already considered in the previous section, is elaborated to account for the core/shell interface.

If one considers merely the cohesive energy of the two crystal phases, a transformation of the crystal structure during the shell formation is expected, from the hexagonal structure of the wurtzitic core template to the cubic symmetry of the lowest-energy crystal phase of Si or Ge. Besides the hidden kinetic barriers involved in the transition mechanism, two important energy costs hinder the crystal phase transformation. Firstly, the surface energy of the sidewalls of the NWs pushes towards the stabilization of the *2H* crystal phase, as discussed

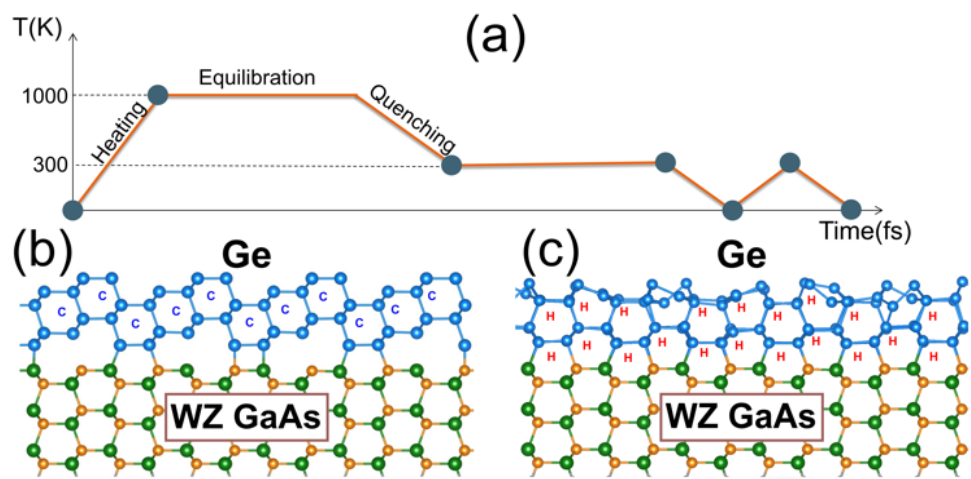

**Figure 6** Atomistic models of the interface between wurtzitic GaAs and *3C*-Ge. On the left side the atomistic model before the simulations, while on the right the structure after the *ai*-MD simulations. The top sketch illustrates the *ai*-MD simulation steps: the system has been heated-up to 1000 K, then equilibrated for few ps, cooled-down to about room temperature and re-equilibrated again, before reaching the temperature of 0 K. Few steps of heating-up and cooling-down have been also performed, from 0 to 300 K and vice versa.

previously. Then, if the crystal transformation occurs, a boundary between the two crystal structures

(hexagonal core and cubic shell) will be formed. While the core/shell interface would be coherent in the case of a *2H* shell, an incoherent boundary should be formed in the case of the *3C* shell. Thus, another important energy term has to be included in favor of the *2H* crystal in the total energy balance. Before modelling an ideal boundary between the two crystals, we perform *ab-initio* molecular dynamics (*ai*-MD) simulations with a rudimental initial model of the interface between wurtzitic GaAs and *3C*-Ge. This slab model is made up of a few the $(10\bar{1}0)$ oriented GaAs layers with Ga and As atoms fixed at their bulk-like position and using the calculated lattice constants of bulk GaAs. On top of the GaAs layers, three Ge layers as-cut from a *3C* crystal with [112] orientation are placed, as illustrated in Fig. 6(b). The Ge atoms are free to rearrange during the *ai*-MD simulation. The *ai*-MD steps are illustrated in the sketch of Fig. 6(a): after initial heating of the system up to a temperature of 1000 K, breaking the initial configuration of the Ge layers, the system is equilibrated for a few ps and then quenched down to room temperature. Then, another quenching step is performed down to 0 K temperature. At this stage, we note recrystallization of the Ge resembling the *2H* crystal. However, several heating and quenching steps from 0 K to 300 K have been necessary to obtain the configuration illustrated in Fig. 6(c), showing the formation of perfect *2H*-Ge layers on GaAs.

This *ai*-MD simulation supports the results presented in the previous section, indeed predicting higher stability of the *2H* crystal phase for such a thin layer of Ge. But the central question on the contribution of the interface of the core/shell NWs to the stability of the *2H* crystal is still open: how much does the need to create a proper interface boundary between the two phases affect the critical thickness for the transformation of the shell into a cubic one? Unfortunately, due to the high computational cost of the *ai*-MD simulation, it is not possible to repeat the simulation with a thicker Ge layer, approaching the nm scale. Even if this were possible, the formation of the interface boundary during the MD simulations would be not trivial both because of the kinetic barrier and the very limited simulation time of the *ai*-MD simulation. However, the radial $(10\bar{1}0)$ hexagonal growth fronts could be truly converted into a cubic (112) oriented one if an incoherent twin boundary is formed. It has been recently observed in high-resolution transmission electron microscopy (HRTEM) real-time experiments on III-V NWs that the transformation mechanism between the WZ and ZB phases occurs via a collective glide of a set of three Shockley partial dislocations located on every two (111) cubic planes [47]. When the three dislocations come close together, a complex is formed, which is periodically repeated with the incoherent boundary. In particular, when the complex is composed by a 90° Shockley partial dislocation inserted between two 30° with opposite screw components (as suggested in Ref. [47]), the relative total Burgers vector is equal to zero, having two important consequences: firstly, the complex does not involve a distortion in the cubic layer (ignoring the local strain field around each partial core) resulting in the lower possible energy cost of the interface; secondly, the three partial dislocations attract each other, making the complex compact and sharp.

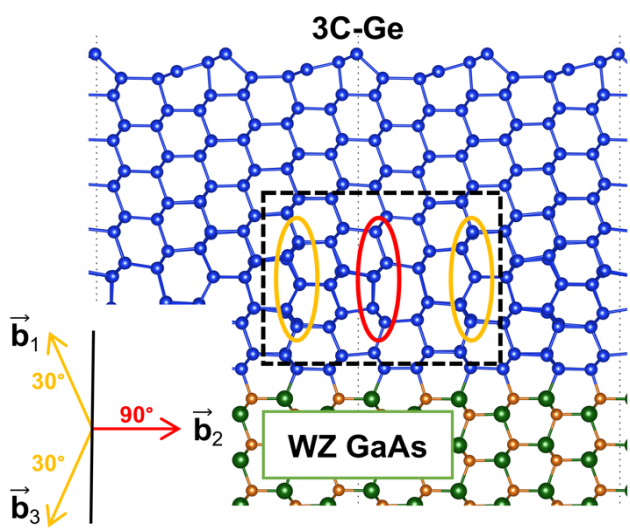

**Figure 7** Atomistic model of the interface between wurtzitic GaAs and *3C*-Ge (a GaP/Si counterpart has been also modeled). The yellow ovals highlight 30° partial dislocations, while the red one indicates the 90° partial. The rectangle with dashed line shows the periodic 6 layers forming the boundary. The sketch on the right illustrates the three Burgers vectors $\vec{b}_1 = 1/6\,[11\bar{2}]$, $\vec{b}_2 = 1/6\,[\bar{2}11]$, , $\vec{b}_3 = 1/6\,[1\bar{2}1]$, , referred to the cubic lattice. Their colors are coherent with those of the ovals, while the black line identifies the dislocation line.

To quantify the interface cost, the described set of Shockley partial dislocation has been inserted in the interface model, as illustrated in Fig. 7. An estimate of the interface energy can be obtained by:

$$E_{int} = \frac{(E_{slab}^c - \mu_c N_c - \gamma_{112}A) - (E_{slab}^h - \mu_h N_h - \gamma_{10\bar{1}0}A)}{A} \quad (3).$$

$E_{slab}^c$ is the total energy of the slab shown in Fig. 7 and $E_{slab}^h$ is the total energy calculated for the same slab but with *2H*-Si or Ge on top, thus forming a coherent interface with the GaP or GaAs substrate. The number of Si or Ge atoms in the slab with *3C* or *2H* crystal is indicated by $N_c$ and $N_h$, respectively. A is the surface area, obviously identical for the two slabs including the *3C* or *2H* crystals.

According to Eq. (3), the calculated interface energy is 66.43 meV/Å² for Si and 47.57 meV/Å² for Ge. If this energy, evaluated for the specific lateral surface of the core, is added to the energy of the *3C* NWs, as expressed in Eq. (2), one can re-calculate the crossover between the formation energy of *3C* or *2H* core/shell NWs. The estimated critical thickness of the *2H* shells on top of GaP or GaAs cores are reported in Table 2, for core radiui of 17.5 or 87.5 nm.

**Table 2** Critical thickness of the 2H Si or Ge shell grown on GaP or GaAs NWs with radii of 17.5 or 87.5 nm.

| Core radius | 17.5 nm | 87.5 nm |
|---|---|---|
| GaP/Si NW | 13.9 nm | 16.0 nm |
| GaAs/Ge NW | 7.6 nm | 8.3 nm |

These two core radii are chosen because they have been experimentally used in the synthesis of *2H*-SiGe, recently reported by the Bakker's group [12]. The two quite different sizes of the cores also allow drawing one interesting conclusion, in view of possible applications. Despite the increase in the critical shell thickness for the larger core is not dramatic, as provided by the larger surfaces/interfaces, the actual volume of the shell increases in a sensible way. As far as no strain is provided by a lattice misfit between the core and the shell, this means that larger cores are more suitable to obtain emitters with larger efficiency.

Note also that the critical thickness for the Si shells is about two times the Ge counterpart, confirming the correlation between Si and Ge NWs found in the previous section, for bare NWs. It is also important to remark the difference in size between the critical radius found before (no core/shell) and the shell thickness in case of core/shell NWs, with the latter ones being about 2-3 times larger. This is an evident indication of the helpfulness of core/shell NWs in the growth of the *2H* Si and Ge crystals. The role of the core as a template for the shell is particularly evidenced: exposing specific crystal facets the sidewalls of the core facilitate the crystallization of the *2H* crystal phase while hindering the growth of *3C* crystals because an incoherent crystal boundary should be also formed. The formation of the incoherent boundaries is an aspect that should be carefully analyzed when the critical thickness calculated here is compared to the shell thickness obtained experimentally. Indeed, while our maximum predicted thickness for the *2H*-Si or Ge shell is in the range of 8-16 nm, the *2H*-NWs shells grown experimentally, particularly by the Bakker's group, can reach few hundred nanometers. But it should be stressed that our model considers the worst-case scenario for the *2H* crystal formation, by comparing the coherent core/shell interface with the ideal and lowest energy incoherent boundary of the WZ/*3C* interface. The model is also simplified because it considers only the thermodynamic driving forces, while the kinetic barrier in the formation of the incoherent boundary is not accounted for. Nevertheless, our results demonstrate the effectiveness of these thermodynamic driving forces in the formation of *2H*-Si and Ge NWs, particularly for core/shell structures, up to several nanometers in thickness. Actually, they are not at all in contrast with the hundreds of nanometers shell thicknesses obtained experimentally. In fact, during the CVD process, the shell growth front advances radially, and our results show that in the first few nanometers, the formation of a metastable *2H* crystal shell is thermodynamically favored over a *3C* shell if specific sidewall facets of the core NWs are used as a template. Then, when the critical thickness is reached, even if a whole *3C* shell becomes thermodynamically favored over the *2H* one, a crystal phase transition of the already formed *2H* shell should occur to effectively see the formation of the *3C* shell and have a total energy gain. But this is practically quite improbable because a crystal phase transition of the Si or Ge shell at the temperatures used in the CVD growth (below 1000 K) is unlikely. Finally, this aspect may be considered as an important advantage of the CVD core/shell growth over the VLS method in the synthesis of *2H*-Si and Ge crystals.

### 3.Conclusion

The diameter dependent stability of *2H*-Si and Ge NWs, including the case of core/shell structures, has been theoretically investigated. By performing atomistic simulations based on the density functional theory, and also exploiting generalized interatomic potential to extend the simulation scale, the formation energy of *2H* and *3C*-Si and Ge NWs has been calculated as a function of the NW radius. The results reveal the higher stability of the *2H* crystals for very small NWs. Crucially, this is due to the lower surface energy of the facets of *2H* NWs, as compared to that of *3C* NWs. This thermodynamic driving force is effective just up to a few nanometers of the NW radius, where a crossover between the formation energy of *2H* and *3C* NWs is expected. The crossover points are indeed found and confirmed by two simulation methods. Thus, critical radii of about 6 and 3 nm are predicted for the *2H*-Si and Ge NWs, respectively. This explains the rare successes in growing *2H*-Si NWs by the VLS method, and the very small radii obtained.

Models of core/shell NWs and including the incoherent boundary between the wurtzitic core and the *3C*-Si or Ge shell have been also considered. The results highlight the fundamental role of the core as a template for the hexagonal shell when specific facets of the WZ crystals are exploited. The presence of the core/shell interface enlarge the thermodynamic stability range of the thickness of the *2H* shells and is at the base of a proposed mechanism elucidating the successful radial growth of nanowires in the metastable phase, up to shell thickness of a few hundred nanometers.

### 4.Method

The DFT simulations were performed within the generalized gradient approximation (GGA) [48]. Projected augmented wave (PAW) pseudopotentials [49,50] were employed, as implemented in the VASP code [51,52]. The energy cutoff was set to 450 eV and a (8×8×4) and (8×8×8) k-point mesh was used for the unit cell of the bulk *2H* and *3C* crystal, respectively. They were adjusted for the slab calculations accordingly to the ratio of their size and the bulk unit cells, but always allowing a convergence of the total energy of the systems below 10 meV. To obtain surface energies (listed in Tab 1) the total energy of symmetric Si or Ge slabs were calculated with periodic boundary conditions in all directions except for the direction perpendicular to the surface investigated. All the structures were relaxed until the forces on all atoms were less than 10 meV/Å. The energy difference of the *2H*-NWs and a *3C* bulk crystal having the same number of atoms was calculated as:

$$\Delta E_h = \frac{(\mu_h - \mu_c) \cdot N + \gamma_{10\bar{1}0} \cdot L_{Surf}}{N} \qquad (4).$$

N is the number of atoms contained in the hexagonal prism and used for modelling the NW, and $L_{Surf}$ its lateral surface. The first term of the sum at the numerator of Eq. 4 represents the bulk-like energy, while the second term is clearly the surface contribution. The first term is obviously zero when a similar expression of the energy difference is formulated for the *3C* NWs:

$$\Delta E_c = \frac{\gamma_{112} \cdot L_{Surf}}{N} \qquad (5).$$

To plot these energy differences as functions of the NW radius ($r_{nw}$), one needs to rewrite the $L_{surf}$ and N in Eqs. 4 and 5 in terms of the radius. This can be done by expressing $L_{surf}$ as a function of the NW volume and finally N, by exploiting the atomic volume ($Vol_{atm}$) of the cubic or hexagonal phase:

$$L_{Surf} = 6h_{nw}\sqrt{Vol_{nw}\frac{2}{h_{nw}3\sqrt{3}}} = 6h_{nw}\sqrt{Vol_{atm}N\frac{2}{h_{nw}3\sqrt{3}}} \qquad (6).$$

Here, the NW radius is written as:

$$r_{nw} = \sqrt{Vol_{atm}N\frac{2}{h_{nw}3\sqrt{3}}} \qquad (7),$$

and an expression for the number of atoms was derived from Eq. 7 as:

$$N = \frac{r_{nw}^2 \, h_{nw}}{Vol_{atm}}\frac{3\sqrt{3}}{2} \qquad (8).$$

The relations between the number of atoms, total volume of the NW and atomic volume of the bulk crystal are quite precise, except for small NWs, particularly when the $r_{nw}$ approaches the lattice parameter of the bulk crystal. A correction term for $r_{nw}$ has been estimated to roughly account for this error, and it is about 0.16 nm for Si NWs and 0.15 nm for Ge NWs. These corrections have been obtained by comparing the radii of the NWs as measured from their atomistic models, with those expected from Eq. 7.

The final expressions corresponding to Eqs. 4 and 5 and containing the dependence on the NW radius are stated in the Eqs. 1 and 2, with specific atomic volume for the two equations, being it slightly different for the *2H* and *3C* phase.

For the *ai*-MD simulations, the same energy cut-off was kept for but using only the gamma point for the interface slab calculations. A canonical ensemble (N, V, T) with Nosé-Hoover thermostat was used during the simulations, with a time step of 1 fs. The heating and quenching ramps were set in order to get 5 simulation steps per K of temperature variation. All the interface slab models employed Hydrogen (or pseudo-Hydrogen) atoms to saturate the terminating bonds of the bottom side of the slabs, to mimic an infinite bulk substrate. But, to avoid interactions between the periodic replica in the non-periodic (out-of-plane) direction, a 15 Å thick vacuum layer was used.

Classical MD simulations were performed using the Large-scale Atomic/Molecular Massively Parallel Simulator (LAMMPS) code [53]. To describe the interaction between the Si atoms, the Csányi's potential [44] was used. This MD potential enables to perform simulations with quantum mechanical accuracy by the realization of an efficient interpolation scheme for the potential energy landscape between preliminarily determined values at different points of atomic configuration space, applying a regression method known as the Gaussian process. Periodic boundary conditions were set in the NW growth direction, i. e. [111] for 3C-Si and [0001] for 2H-Si NWs.

The 2H and 3C-Si nanowire were simulated using segments analogous to those shown in Fig. 4, with radii ranging from 0.2 to 18 nm, performing MD simulations in the canonical ensemble (N, V, T), using a Nosé-Hoover thermostat regime. To induce the surface reconstruction, NW segments were annealed at 300 K during 1 ps of simulated time followed by a decrease of temperature down to 0.01 K during another 1 ps. The time step of simulations was 1 fs based on the energy conservation in the course of preliminary simulation runs. After energy minimization procedure with the conjugate gradient algorithm, the total energy of the simulation cell was determined.

## Acknowledgements

The CINECA consortium is acknowledged for the availability of high-performance computing resources, granted under the ISCRA initiative.

D.B. Migas acknowledges the partial financial support of the "Improving of the Competitiveness" Program of the National Research Nuclear University MEPhI – Moscow Engineering Physics Institute.

We are also grateful to Federico Grassi for his work on this topic during his Master Thesis at the University of Milano-Bicocca.